\documentstyle[amsfonts,amssymb,preprint,aps]{revtex}
\tightenlines
\begin{document}

\title{Graph kinematics of  discrete physical
objects: \\
beyond space~- time. I.~General}
\author{V. E. Asribekov}

\address{All~- Russian Institute for Scientific and
Technical Information, VINITI, Moscow 125315, Russia\\
(\rm e-mail:peisv@.viniti.ru)}
\maketitle

\begin{abstract}
The necessity of an introduction of discrete physical objects in
physics conception is analysed taking into consideration an optimum
stage for postulating of some like objects in microworld as well as
in macroworld including the new ``physical graph'' as a discrete
microobject and carrying out its analogy with ``Kirchhoff's laws
graph'' for an electric network as a prototype of discrete
macroobject which correspond both to discrete sets of trees~---
root trees (for microobjects) or skeleton trees (for macro
networks).
The transitions are found connecting the usual
$S$-matrix theory with Feynman integrals and Feynman diagrams and
the new physical graph kinematics formalism which uses the natural
root trees basis for the treatment of the structure of an arbitrary
complicated physical microobject with a specific ``graph
microgeometry''~--- beyond space-time consideration. Accordingly
to the QCD results the proton (nucleon) mass is determined in
terms of the root trees number $T_{v=11}$=1842 which corresponds to
$v$=11 physical graph vertices. It is supposed that by means of a
double- and a triple-splitting of the root trees numbers could be
estimated the masses of the various series of another
microobjects.
\end{abstract}


\section*{1 Introduction}

In our days the continuous physical objects fashion especially based on
their field-theoretical description is very strong. In this connection it is
really hard for any other point of view to gain a hearing.

Of course, it is naturally that in case of the macroobjects it is not necessary
obviously to introduce some alternatives for this standard approach within a
customary ``external'' geometry. Nevertheless, beginning from the most
elementary viewpoint for describing atomic events we could not picture how
the jump from one electronic orbit to another took place and we just had to
accept it as a kind of discontinuity.

The following evolution of a picture of the nature towards the nuclear and
subnuclear microobjects picture leads us to more open discontinuities and at
last to the evident discrete physical objects perhaps with the proper
``internal'' geometry corresponding to their inner probably discrete structure.
It it notable that a well-known field-theoretical problem of the divergence
difficulties at the small distances for now structural physical microobjects
eliminates automatically what could be considered as an essential result
in the area of discrete objects physics.

In parts I--III of paper we describe a possible transition to the
some representation of discrete physical objects using the graph
theory formalism.

It is important to have in view that in general a discrete
mathematics as well as the physical theories and models with any
discrete structural objects are not derived from or reduced to the
continuous mathematics and the physical theories and models with
corresponding continuous objects. Therefore the real appearance in
physics of various discrete physical objects can be made solely by
the introduction of an adequate mathematical postulate without some
additional justifications but using any typical ``discrete-like''
results from an initial quasi-continuous physical theory;
nevertheless the continuous theories altogether are not excluded
from a following consideration. For this purpose however it is
necessarily to find a definite stage in continuous physics
development setting up the insufficiency of its continuous
theory. Inasmuch as the above-mentioned difficulties arise in
phenomena involving very small distances (or very high energies) we
may choose as a such stage the transition to microworld. In this
connection part~I includes Feynman diagram technique within
{\it S}-matrix theory for microobjects. Taking into account a
non-equivalence of some following singularities--analysed diagram
technique (post--Feynman), based on the stable
microobjects only, to a perturbation theory we postulate a new
derivative physical graph formalism as a first step to discrete
microobject. Owing to an analogy of such physical graph with
Kirchhoff's laws graph for an electric network there exist various
discrete physical objects which may be presented through the
discrete sets of graphs---skeleton or root trees, beyond usual
space-time. In part~II a proposed graph formalism is applied to
the calculation of some qualitative and numerical characteristics
for different microobjects without using of the continuous theories
formalism (QED, QCD, etc.) itself but only its results. And
part~III is devoted to a possible realization of the
Heisenberg---Dyson's two-layer physics.

\subsection*{1.1 Heisenberg's $S$-matrix \\
of 1943 and its state vectors basis \\
in the Hilbert space}

It is known that proposed by Heisenberg in 1943 the $S$-matrix
contains the only physically measurable quantities and supposes the
existence of a corresponding Hilbert space with the positive metric
and the possibility to construct the complete Hilbert space basis.
The $S$-matrix must be unitary and have so much analyticity that
it represents what observed as causality; it also must have an
invariance for the Lorentz group and for {\it TCP}, an approximate
invariance for the isospin group, and so on.

Since 1948--49 the Feynman version of QED became the
prototype of what is now called $S$-matrix theory which gave
directly the rules for calculating $S$-matrix elements by means of
Feynman integrals and corresponding Feynman diagrams. It is
important the Feynman theory is a pure physical object theory, and
the Feynman diagram describes any elementary process naively as a
propagation of physical object from one vertex to another along a
connected line.

\subsection*{1.2 ``Singularities matrix''
for Feynman integrals \\
and networks of the  new physical graphs \\
in momentum space}

A consistent analysis of all possible singularities of Feynman
integral as a key quantity of the $S$-matrix formalism in Hilbert
space (see for example, Ref. [1])

$$\displaystyle
\int\prod\limits^{l}_{r}d^{4}k_{r}\prod\limits^{n}_{s}f^{-n}\delta\!\left(\sum
\alpha -1\right)d\alpha_{s}, \eqno(1)$$

$$\displaystyle
f=\sum\limits^{n}_{i}\alpha_{i}\,(m_{i}^{2}-q_{i}^{2}), \eqno(2)$$

determining the contribution of an arbitrary Feynman diagram with
$N$ external 4-momenta $\displaystyle p_{j}$ $(j=1,\,2,...,\,N)$,
$n$ internal 4-momenta $\displaystyle q_{i}$ $(i=1,\,2,...,\,n)$
and $l$ loop momenta $\displaystyle k_{r}$ ($r=1,\,2,...,\,l)$ can
in principle be performed by solving a set of corresponding $v$
laws of conservation of  4-momenta in $N$ external (algebraic sum
over [$j$]) and $v-N$ internal (algebraic sum over ($i$)) vertices
and $l$ Landau extremal independent loop equations (algebraic sum
over $<r>$) in fact already for a new derivative (post--Feynmanian)
physical graph
with the same characteristics

$$ \ \displaystyle\sum\limits_{[j]}\epsilon q=p_{j}; \
\ \ j=1,\,2,...,\,N,
\eqno (3)$$
$$\rule{14dd}{0dd}\displaystyle\sum\limits_{(i)}\epsilon q=0; \ \
\ i=1,\,2,...,\,v-N, \eqno (4)$$
$$\displaystyle\sum\limits_{<r>}\alpha q=0; \ \ \ r=1,\,2,...,\,l.
\eqno (5)$$

Actually the set from (3) and (4) contains only
$v-1$ independent equations, since one of the equations corresponds to
the law of conservation for the external 4-momenta $\displaystyle
p_{j}$ $$\displaystyle\sum\limits^{N}_{j}\epsilon p=0 \eqno(6)$$

(everywhere $\displaystyle\epsilon =0,\, \pm 1).$

Therefore the total number of independent equations (3), (4) and
(5) is equal precisely to the number of the unknown internal
4-momenta $q$
$$\displaystyle [N+(v-N)]-1+l=v-1+l=n.$$

The rank of the obtained square matrix of coefficients from
equations (3), (4) and (5)~--- so-called ``singularities matrix''
for the Feynman integral (1)
$$\displaystyle \mathop{\bf M}_{(n\times n)}=\left\{
\begin{array}{c}
\displaystyle\mathop{{\bf I}\,(\epsilon )}_{(v-1\times n)} \\
\displaystyle\mathop{{\bf A}\,(\alpha)}_{(l\times n)} \\
\end{array}\right\}\eqno (7)$$

as can be shown is also equal to $n$.

This composite matrix $\displaystyle\mathop{\bf M}_{(n\times n)}$
contains the incidence matrix
$\displaystyle\mathop{\bf I}_{(v-1\times
n)}\!\!\!\!(\epsilon_{ij})$ and the independent loop  matrix
$$\displaystyle \mathop{\bf A}_{(l\times
n)}\!\!(\alpha)=\mathop{\bf C}_{(l\times n)}\!\!(\epsilon_{ij})
\mathop{{\bf D}_{n}}_{(n\times n)}\!\!(\alpha )$$ (where
$\displaystyle\epsilon_{ij}=0,\, \pm 1)$; here
$\displaystyle\mathop{\bf C}_{(l\times n)}\!(\epsilon_{ij})$ is an
usual cyclic (circuit) matrix and $\displaystyle
\mathop{{\bf D}_{n}}_{(n\times n)}\!\!(\alpha)$~--- diagonal matrix

$$\displaystyle
{\bf D}_{n}\,(\alpha)=
\displaystyle \left(
{\arraycolsep=1.5dd
\begin{array}{cccc}
\alpha_1 & & &                        \\
&\alpha_2  & &                                                  \\
& & \ddots{} &                                           \\
& & & \alpha_n                              \\
\end{array}}\right).$$

It is known (see Ref.~[2]) that
$\displaystyle \mathop{\bf C}_{(l\times n)}\times\mathop{{\bf
I}^{\bf T}}_{(n\times v-1)}\equiv 0.$

The rest of Landau extremal equations for internal 4-momenta

$$\displaystyle q^{2}_{i}=m_{i}^{2}; \ \ \ i=1,\,2,...,\,n
\eqno(8)$$

together with the initial conditions for external 4\--mo\-menta

$$\displaystyle p_{j}^{2}=M_{j}^{2}; \ \ \ j=1,\,2,...,\ N,
\eqno(9)$$

where $\displaystyle m_{i}$ and $\displaystyle M_{j}$ are the
masses of internal and external lines in a new derivative physical
graph, set up that all 4-momenta~--- an internal $\displaystyle
q_{i}$ as well as an external $\displaystyle p_{j}$~--- can be
located according to Landau (see Ref.~[3]) factually on the mass
shell.

Thus the physical graph includes in fact only the real stable
physical objects and the matrix equation

$$\displaystyle \mathop{\bf M}_{(n\times
n)}\mathop{(Q_{n})}_{(n\times 1)}=\mathop{(P_{n})}_{(n\times
1)}\eqno (10)$$

where by means of $\displaystyle\mathop{(Q_{n})}_{(n\times 1)}$ and
$\displaystyle\mathop{(P_{n})}_{(n\times 1)}$ are denoted the
column vectors
$$\displaystyle \mathop{(Q_{n})}_{(n\times
1)}=\left(\begin{array}{c}
q_{1}                                                       \\
q_{2}                                                        \\
q_{3}                                                         \\
q_{4}                                                         \\
\vdots{}                                                      \\
q_{n-1}                                                        \\
q_{n}
\end{array}\right),\ \mathop{(P_{n})}_{(n\times
1)}=\left(\begin{array}{c}
p_{1}                                                         \\
p_{2}                                                          \\
\vdots{}                                                       \\
p_{N-1}                                                        \\
0                                                               \\
\vdots{}                                                        \\
0                                                              \\
\end{array}\right) \eqno(11)$$

may serve as a starting point for the following consistent
consideration of the networks of these physical graphs in momentum
space. Although it is necessarily to note that the direct solution of
the full set of algebraic equations (3)--(6), (8), (9)~--- for a
separated as well as for a complicated network of such physical
graphs~--- is enough complex and labor-intensive but quite
feasible. The fact is that the results may be classified on the
base of the forms of construction of the special $l,\,v$~---
sequences of physical graphs, namely the ladder and the parquet
$l,\,v$~--- sequences.

\section*{2 Kirchhoff's laws matrix for \\[2dd]
an electric network
analogy of 1847}

In agreement with the classical Kirchhoff's work of 1847 (see Ref.
[4]) initiated a development of the graph theory (especially in
part of the particular graphs~--- the trees) the investigation of an
arbitrary electric network with $n$ wires is carried out by means
of two laws:

$$\!\!\displaystyle \sum\limits^{}_{[j]}\epsilon J=J^{*}_{j}; \ \
\ \ j=1,2,\ldots{} ,N,\eqno(12a)$$

$$\displaystyle \sum\limits^{}_{(i)}\epsilon J=0;\ \ i=1,2,\ldots{}
,v-N,\eqno(12b)$$

$$\displaystyle \sum\limits^{}_{<r>}J\omega=\sum\limits^{}_{<r>}{\cal E};\ \ r=1,2,\ldots{} ,l.\eqno(13)$$

The first, current law for $N$ external (algebraic sum over [$j$] in
(12a)) and for $v-N$ internal (algebraic sum over ($i$) in (12b))
vertices of the electric network states that the algebraic sum of
the currents $J$ flowing through all the network wires that meet at a
vertex is the external current $J^{*}$ or zero. The second, voltage
law for $l$ circuits (loops) states that the algebraic sum of the
electromotive forces ${\cal E}$ within any closed circuit is equal
to the algebraic sum of the products of the currents $J$ and the
resistances $\omega$ in the various portions of the circuit.

The Kirchhoff's laws (12a), (12b), (13) for an electric network
including only the real measurable quantities $J,{\cal E}$ (and
also $\omega$) are formally the same as the linear equations
(3)--(5) for an identical topologically physical graph with real
objects quantities $q,\ p$. The electric circuit analogy still with
the suitable Feynman diagram has been carried by Bjorken, T.~T.~Wu
and Boiling earlier in 1959--64 (see Ref. [1]). It is easy to show
(see for example, Ref. [5]) however that the number of independent
equations among (12a)--(12b) as in the case of set (3) and (4) is
equal to $v-1$ and therefore the rank of the resulting square
matrix of coefficients

$$\displaystyle \mathop{\bf M}_{(n\times
n)}^{}=\left\{\begin{array}{c}
\displaystyle\mathop{{\bf I}(\epsilon)}_{(v-1\times n)} \\
\displaystyle\mathop{{\bf C}(\epsilon)}_{(l\times
n)}^{}\displaystyle\mathop{{\bf D}_{n}(\omega)}_{(n\times n)}^{}

\end{array}\right\}\eqno(14)$$

is exactly corresponded to the number of unknown components of the
column vector $(\displaystyle J_{n})$ i.~e. to $n$.

In this way the soluble matrix equation for a typical Kirchhoff's
matrix (14) may be written in the obvious form

$$\displaystyle \left\{\begin{array}{c}
\displaystyle\mathop{{\bf I}(\epsilon)}_{(v-1\times n)} \\
\displaystyle\mathop{{\bf C}(\epsilon)}_{(l\times
n)}^{}\displaystyle\mathop{{\bf D}_{n}(\omega)}_{(n\times n)}^{}
\end{array}\right\}\displaystyle
\mathop{(J_{n})}_{(n\times
1)}^{}=\left(\begin{array}{c}
\displaystyle\mathop{(J^{*}_{v-1})}_{(v-1\times 1)}^{} \\
\displaystyle\mathop{{\bf C}\,(\epsilon)}_{(l\times
n)}^{}\displaystyle\mathop{({\cal E}_{n})}_{(n\times 1)}^{}
\end{array}\right)\eqno(15) $$

where by means $\displaystyle \mathop{(J_{n})}_{(n\times 1)}^{},\
\displaystyle \mathop{(J^{*}_{v-1})}_{(v-1\times 1)}^{}$ and
$\displaystyle \mathop{({\cal E}_{n})}_{(n\times 1)}^{}$ are denoted
the column vectors

$$\displaystyle \mathop{(J_{n})}_{(n\times
1)}^{}=\left(\begin{array}{c}
J_{1}                                                            \\
J_{2}                                                            \\
J_{3}                                                            \\
\vdots                                                            \\
J_{n-1}                                                          \\
J_{n}
\end{array}\right), \
\mathop{(J^{*}_{v-1})}_{(v-1\times
1)}^{}=\left(\begin{array}{c}
J^{*}_{1}                                                        \\
\vdots{}                                                         \\
J^{*}_{N-1}                                                      \\
0                                                                \\
\vdots{}                                                         \\
0
\end{array}\right),$$

$$\displaystyle
\mathop{({\cal E}_{n})}_{(n\times 1)}^{}=\left(\begin{array}{c}
{\cal E}_{1}                                                     \\
{\cal E}_{2}                                                     \\
{\cal E}_{3}                                                     \\
\vdots{}                                                          \\
{\cal E}_{n-1}                                                   \\
{\cal E}_{n}
\end{array}\right).\eqno(16)
$$

\subsection*{2.1 Skeleton trees basis\\
for electric network}

In accordance to Kirchhoff's theorem (see Refs. [2], [5]) every
electric network can be substituted by a corresponding graph with
the same number of vertices $v$. The solution of the equations (12)
and (13) for this adequate graph may be presented through the set
of skeleton $v$-trees. The maximum set of independent
skeleton $v$-trees for the corresponding full $K_{v}$-graph,
with $\displaystyle \left(v\atop 2\right)$ lines, forms the skeleton
trees basis for electric network. By using of this basis can be
constructed the all possible solutions for various concrete
networks (lots of such electric network examples described
in Ref. [5]).

\subsection*{2.2 Kirchhoff\---Maxwell topological analysis\\
of electric networks}

Two-layer structure of the typical Kirchhoff's matrix (14) (or
analogous matrix (7)) in a general matrix equation (15) (or in
corresponding matrix equation (10)) shows a separation of the
functions of the matrix operator (14) (or (7)) acting, firstly,
upon the ``internal geometry'' of electric network by means of the
incidence matrix $\displaystyle\mathop {{\bf
I}(\epsilon)}_{(v-1\times n)}$ (``upper layer'') and, secondly, upon
the topologically other equilibrium conditions along the circuits
(loops) of electric network by means of the loop matrix
$\displaystyle\mathop{{\bf C}(\epsilon)}_{(l\times n)}$
\,$\displaystyle\mathop{{\bf D}_{n}(\omega)}_{(n\times n)}$ (``under
layer'') within the framework of the same skeleton trees basis. In
this connection it is important to note that into the structure of
the incidence matrix
$\displaystyle\mathop{{\bf I}(\epsilon)}_{(v-1\times n)}$ is embedded
the ``internal geometry'' of real electric network including the
natural skeleton trees basis factually as a creation of the
specific ``graph geometry'' beyond the space-time consideration.

By introducing of ``circuit currents'' of
Helmholtz\---\-Maxwell\- instead of ``branch (wire) currents'' of
Kirchhoff can be performed formally the analogous concrete
calculations on the base of corresponding topologically equivalent
Maxwell rules (see Ref. [5]) within the same ``graph geometry''.

Obviously, the last formalism of the Helmholtz\---\-Max\-well ``circuit
currents'' is like to the formalism of
``($p,\,k$)\--dia\-grams'' (see Ref. [6]) which may be used as an
adequate tool for the analysis of the physical graphs from section
1.~2. However this problem is beyond the task of a given paper.

\section*{3 ``Microgeometry'' inside\\
of the physical objects\\
considering within the framework \\
of the input\---output scheme}

Heisenberg's theory of the $S$-matrix connects the input and
output of a scattering experiment without seeking to give a
localized description of the intervening events including the inner
structure of a propagated physical object.

Introducing a full set of the external 4-momenta $p_{j}$ ($j$=1, 2,
\ldots{}, $N$) simultaneously for $N_{1}$ ingoing as well as for
$N_{2}$ outgoing physical objects $(N_{1}+N_{2}=N)$ we can obtain
the solution of the equations (3)\---(6), together with the mass
shell conditions (8)\---(9), in the terms of independent kinematic
invariants $s_{i}=p_{i}^{2}$, $s_{ik}=(p_{i}+p_{k})^{2}$,
$s_{ikl}=(p_{i}+p_{k}+p_{l})^{2}$,
$s_{iklm}=(p_{i}+p_{k}+p_{l}+p_{m})^{2}$, etc. connecting with each
other by kinematic and geometrical conditions in the
4-dimensional momentum space (see Refs. [7--10]). Thus we should
be presented the whole physical picture for the real scattering
process where an inner structure of the participating physical
objects is factually omitted and the latter should be introduced only by
insertion of the separate full graph with $v$ vertices $K_{v}$ as a
some complicated vertex-fragment for concrete physical object.

\subsection*{3.1 Root trees basis for\\
physical graph of microobject}

Within the framework of the input\---output scheme for the full
physical graph with inserted complicated $K_{v}$-vertex~, as a
physical microobject vertex-fragment in the general scheme~, it is to be
determined the special separate vertex~--- the root of tree which
coincides with input vertex, and the other part of physical
graph is to be arranged in hierarchical order, creating a path to
output vertex (vertices).

Thus the solution of the corresponding linear equations (3)--(5)
for this new hierarchical physical graph may be presented already
through the set of root $v$-trees as against of skeleton
$v$-trees for electric network in section 2.1.

If we consider the full physical graph as an unique $K_{v}$-vertex
in the input\---output scheme then may be formulated the problem of
the discrete physical object structure. The maximum set of
independent root $v$-trees for corresponding full $K_{v}$-graph,
with the separate root-vertex and $\displaystyle \left(
v\atop 2\right)$ lines, forms the natural root trees basis for the
hierarchical physical graph, i.~e. a set of paths between the
root-vertex, as an initial point of input, and the final points of
output.

Returning to the two-layer structure of the matrix (7) in a general
matrix equation (10) already for the discrete physical microobjects
in the $S$-matrix input\---output scheme we note again that the
incidence matrix $\displaystyle \mathop{\bf I\,}_{(v-1\times n)}\!\!
\!\!\!\!\!\!(\epsilon)$ (``upper layer'') acts upon the
``internal microgeometry'' of the physical microobjects within the
framework of the natural fixed root trees basis. Therefore the
incidence matrix $\displaystyle \mathop{\bf I}_{(v-1\times
n)}\!\!\!\!\!\! \!(\epsilon)$ contains a specific ``graph
microgeometry'' beyond space-time approach.

\subsection*{3.2 On the proton mass}

The fact is that the above-described initial $S$-matrix theory
overgrew practically in the specific physical graph kinematics
formalism allowing to realize the discrete physical microobject
calculations without use of the traditional space-time treatment.

We start with the consideration of the proton mass problem within
the framework of the results of QCD-theory [11]. It is shown
earlier (in 3. 1) that ``internal microgeometry'' of the discrete
physical microobjects reflects their inner structure and therefore
is to correspond to the vertex many-point Green function
(without free  tails).  These points are responsible for determination
of the number
of vertices of the root trees filling in graph microgeometry of the
discrete physical microobjects (in a Riemann [15] sense).

If, as usually supposed, the full proton mass is concentrated into an
internal gluon field (Refs. [12--13])~--- a case of the pure
gluodynamics~--- then from the Gell-Mann\---Low function (Ref. [14])
$$\displaystyle \beta (g^{2}) =-b \frac{g^{4}}{16 \pi^{2}} +{\rm O}
\left(\frac{g^{2}}{4 \pi}\right)                          $$

where $g$~--- coupling constant, we have a dimensionless constant $b$=11
characterized the number of vertices of the corresponding root
trees. For setting up the graph equivalent to the electron mass
$m_{e}$ we choice the corresponding simplest case of the root tree
graph with $v$=2 vertices inasmuch as the number of a such root
tree $T_{v=2} =1$ (the case $v$=1 is the trivial graph --- an
isolate vertex). Therefore the number of the root trees with $v$=11
vertices $T_{v=11}=1842$ (see Ref. [2]) determines for the proton
mass

$$M_{p}=1842 m_{e}.$$

In the standard QCD-model with the number of quark flavours $n_{f}$
we have  a dimensionless constant $b=11-2/3 \,  n_{f}$ what is
responsible for the pion mass at
$n_{f}$=3: \ $T_{v=9}$=286 (see Ref. [2]) or

$$m_{\pi}=286~m_{e}.$$

\subsection*{3.3 On the ``double~- and triple~- splitting''\\
mechanism in the physics and the biology hierarchy}

The investigation of the inner structure of the discrete physical
microobjects by using of the analysis of a specific ``graph
microgeometry'' may be continued on the base of the all kinds of
root $v$-trees from $v$-sequences (see Table 1) what could permit
to classify the various discrete
\vskip 0dd

{\tabcolsep=3dd
\begin{center}
\newcommand{\Sloppy}{\emergencystretch 3em \tolerance 9999 }
\let\sloppy=\Sloppy
\begin{tabular}{r|r|c}
\multicolumn{3}{c}{{\bf Table 1.} The number of root trees
$T_{v}$ }\\
\multicolumn{3}{c}{and their relations $T_{v+1}$/T$_{v}$
where $v$~--- }\\
\multicolumn{3}{c}{the number of vertices (see Ref. [2]) } \\
\multicolumn{3}{c}{}\\[-5dd]
\hline
\multicolumn{1}{c|}{}       &
\multicolumn{1}{c|}{}       &
\multicolumn{1}{c}{}        \\[-5dd]
\multicolumn{1}{c|}{$v$}       &
\multicolumn{1}{c|}{$T_{v}$}       &
\multicolumn{1}{c}{$T_{v+1}$/$T_{v}$}        \\
\multicolumn{1}{c|}{}       &
\multicolumn{1}{c|}{}       &
\multicolumn{1}{c}{}        \\ [-5dd]
\hline
& & \\
1 & 1 & 1\rule{16dd}{0dd}                  \\
2 & 1 & 2\rule{16dd}{0dd}                  \\
3 & 2 & 2\rule{16dd}{0dd}                  \\
4 & 4 & 2,250                              \\
5 & 9 & \rule{8dd}{0dd}2,22(2)             \\
6 & 20 & 2,400                             \\
7 & 48 & 2,396                             \\
8 & 115 & 2,487                            \\
& & \\[-5dd]
\hline
& & \\ [-5dd]
9 & 286 & 2,514                            \\
10 & 719 & 2,562                           \\
11 & 1842 & 2,587                          \\
& & \\  [-5dd]
\hline
& & \\ [-5dd]
12 & 4766 & 2,620                          \\
13 & 12\,486 & 2,641                       \\
14 & 32\,973& 2,663                        \\
15 & 87\,811& 2,681                        \\
16 & 235\,381& 2,697                       \\
17 & 634\,847& 2,711                       \\
\end{tabular}
\end{center} }

{\tabcolsep=3dd
\begin{center}
\newcommand{\Sloppy}{\emergencystretch 3em \tolerance 9999 }
\let\sloppy=\Sloppy
\begin{tabular}{r|r|c}
\multicolumn{3}{c}{}\\[-5dd]
\hline
\multicolumn{1}{c|}{}       &
\multicolumn{1}{c|}{}       &
\multicolumn{1}{c}{}        \\[-5dd]
\multicolumn{1}{c|}{$v$}       &
\multicolumn{1}{c|}{$T_{v}$}       &
\multicolumn{1}{c}{$T_{v+1}$/$T_{v}$}        \\
\multicolumn{1}{c|}{}       &
\multicolumn{1}{c|}{}       &
\multicolumn{1}{c}{}        \\ [-5dd]
\hline
& & \\
18 & 1\,721\,159& 2,724                    \\
19 & 4\,688\,676& 2,736                    \\
20 & 12\,826\,228& 2,746                   \\
21 & 35\,221\,832& 2,756                   \\
22 & 97\,055\,181& 2,764                   \\
23 & 268\,282\,855& 2,772                  \\
24 & 743\,724\,984& 2,779                  \\
25 & 2\,067\,174\,645& 2,786               \\
26 & 5\,759\,636\,510&                     \\
\end{tabular}
\end{center} }
\vskip 4dd

microobjects and to estimate their
masses. It is easy to see from the Table 1 that $T_{v+1}/T_{v}$
relation is into the interval

\vskip 4dd
\begin{center}
$2\leqslant  T_{v+1}/T_{v}<3$
\end{center}
\vskip 8dd

and therefore we could produce the double- and triple-splitting
operation on the ``mass-values'' of the various $T_{v}$ before the
experimental fitting. In Table 1 we may have conventionally three
zones:

inequality 1$\leqslant T_{v} \leqslant $115 for the stable atomic
and nuclear discrete objects quantities,

inequality 286 $\leqslant T_{v}
\leqslant $1842 for the pion-nucleon discrete objects quantities,

inequality 4766 $\leqslant  T_{v}$ for the heavy physical and the
complicated biological hierarchical objects quantities.

\section*{4 Conclusions}

In conclusion it is important to note that the development of the
graph kinematics formalism for the discrete physical objects
(in a Riemann [15] sense) which has two-layer matrix description
(incidence+loop matrices) with natural ``graph microgeometry''
beyond the traditional space-time consideration leads obviously to
the more general results in the area of the creation of the
Heisenberg\---Dyson's two-layer physics (see for example,
[16]) with the proper ``internal geometry''. In the upper layer of
this scheme we have the formalism for real physical objects, their
momenta, energy, forces, etc. Whereas in
the under layer we have the symbolical fields quantities, such as
field strength, induction, intensity, etc. which may be discovered
only through their energy and forces in the upper layer.

\end{document}